\documentclass[aps,preprint,pra,showpacs]{revtex4}
\usepackage{graphicx}

\def\der#1#2{{\partial #1\over \partial #2}}

\def\be{\begin{equation}}
\def\ee{\end{equation}}
\def\bea{\begin{eqnarray}}
\def\eea{\end{eqnarray}}
\def\bse{\begin{subequations}}
\def\ese{\end{subequations}}
\def\bsea{\begin{subeqnarray}}
\def\esea{\end{subeqnarray}}
\def\({\left (}
\def\){\right )}
\def\[{\left [}
\def\]{\right ]}
\def\<{\left <}
\def\>{\right >}

\begin{document}

\title{Derivation of the Local-Mean Stochastic Quantum Force}

\author{Roumen Tsekov}
\affiliation{Department of Physical Chemistry, University of Sofia, 1164 Sofia, Bulgaria}

\author{Eyal Heifetz}
\affiliation{Department of Geosciences, Tel-Aviv University, Tel-Aviv, Israel}

\author{Eliahu Cohen}
\affiliation{H.H. Wills Physics Laboratory, University of Bristol, Tyndall Avenue, Bristol, BS8 1TL, U.K}

\date{\today}

\begin{abstract}
We regard the non-relativistic Schr\"{o}dinger equation as an ensemble mean representation of the stochastic motion of a single particle in a vacuum, subject to an undefined stochastic quantum force. The local mean of the quantum force is found to be proportional to the third spatial derivative of the probability density function, while its associated pressure is proportional to the second spatial derivative. The latter arises from the single particle diluted gas pressure, and this observation allows to interpret the quantum Bohm potential as the energy required to put a particle in a bath of fluctuating vacuum at constant entropy and volume.

The stochastic force expectation value is zero and is uncorrelated with the particle location, thus does not perform work on average. Nonetheless it is anti-correlated with volume and this anti-correlation leads to an uncertainty relation. We analyze the dynamic Gaussian solution to the Schr\"{o}dinger equation as a simple example for exploring the mean properties of this quantum force. We conclude with a few possible interpretations as to the origins of quantum stochasticity.

\end{abstract}


\maketitle

\section{Introduction}

Indeterminacy seems to be inherent to quantum mechanics. It has been therefore quite appealing to treat the latter like a stochastic theory during the last 70 years. This approach apparently dates back to F$\acute{\text{e}}$nyes \cite{Fey46,Fey52} (criticised in \cite{Nic54}), as well as other authors \cite{Nov51,Tak52,Ker64}. However, it is most recognized through the seminal work of Nelson \cite{Nelson66,Nelson85}, attempting to show that the non-relativistic Schr\"{o}dinger equation (hereafter SE) can describe the ensemble mean dynamics of particles experiencing stochastic Newtonian motion. Stochastic quantum mechanics has been extensively studied and developed since then, e.g. in \cite{Dlp67,Dlp69,Dav79a,Dav79b,Gue81,Gue83} (for recent reviews see \cite{Nel12,Emerg2015}). At first, this approach was based on the assumption that quantum stochastic motion is Brownian, but it was soon understood that quantum stochasticity is conceptually very different from Brownian motion \cite{Emerg2015,Dlp82}.

Here we take an alternative approach. The underlying physical picture is of a massive particle interacting with the vacuum, behaving as some classical-like background medium, so as to cause statistical fluctuations in the position and velocity of the particle. However, we do not assume a-priori any prescribed stochastic mechanism, but rather perform a pure statistical mechanics analysis and then compare it with the Madelung equations \cite{Mad}, studied in our previous works \cite{HC,Entropy,Mott}.

We consider for lucidity the simplest version of the time-dependent SE in 1D, in the absence of external forces:
\be
i\hbar\der{\Psi}{t} = {\hat{p}^2 \over 2m}\Psi =
-{\hbar^2\over 2m}\der{^2 \Psi}{x^2}\, , \hspace{0.5cm}
\Psi({x},t) = \sqrt{\rho}(x,t)e^{iS({x},t)/\hbar}.
\ee
In practice, the probability density function (PDF) $\rho(x,t)$, is evaluated by repeating an identical experiment a large number of times $N$. Hence, in the limit of
$N \rightarrow \infty$ we refer to the PDF in (1) as the ensemble definition:
\be
\rho(x,t) = \lim_{N \rightarrow \infty}\sum_{n=1}^{N}P_n\delta(x-X_n) = \lim_{N \rightarrow \infty}{1\over N}\sum_{n=1}^{N}\delta(x-X_n)\, , \hspace{0.5cm}
\int_{-\infty}^{\infty} \rho(x,t) dx = 1,
\ee
where $X_n(t)$ is the location of the particle in the $n^{th}$ realization at time $t$ and $\delta$ denotes the Dirac delta function. $P_n$ is the statistical weight. Since the experiments are presumed identical and
$\lim_{N \rightarrow \infty}\sum_{n=1}^{N}P_n = 1$, it follows that $P_n = 1/N$.

The questions we ask here are the following. Can SE describe the ensemble mean dynamics of a single particle obeying Newton's second law:
\be
m\ddot{X}_n = F_n,
\ee
where $F$ is a yet undefined stochastic force, whose various realizations $F_n$ are mutually different in general.
Furthermore, if the answer above is affirmative, what information can be obtained regarding the ensemble mean properties under this dynamical law?

We now set off for finding the answers to these questions. Several other routes have been intensively investigated in the past, most notably stochastic quantum mechanics as discussed above and Bohmain mechanics \cite{Bohm,Wya,Holl,Ori}, and indeed several of our results will naturally coincide with those they have found. However, we believe our attempt is unique, starting from a different and very minimal set of statistical definitions, assuming no ontology whatsoever. Our main result will be the derivation of the quantum stochastic force and its properties from basic premises, but as far as we know, the approach in itself is original.

The article is organized as follows. In Section 2 we derive the relevant general properties of the ensemble mean dynamics obeying (3). Next, in Section 3, we write SE as the Madelung equations to formulate the ensemble mean properties it describes. In Section 4 we examine the ensemble mean formulation using the simple example of dynamic Gaussian solutions and finally discuss the results in Section 5.

\section{General properties of ensemble mean dynamics}


We define two averaging operators. The first is the global mean (or expectation value, or just mean):
\be
\<f\> \equiv \lim_{N \rightarrow \infty}{1\over N}\sum_{n=1}^{N} f_n,
\ee
so we can write:
\be
\rho = \< \delta(x-X) \>.
\ee
The second is the local mean:
\be
{\overline f}(x,t) \equiv \lim_{N \rightarrow \infty}\[{\sum_{n=1}^{N} f_n \delta({x}-{X}_n) \over \sum_{n=1}^{N} \delta({x}-{X}_n)}\] =
{\<f \delta(x-X) \>\over \rho}.
\ee
Equations (4) and (6) imply:
\be
\<f\>(t)= \int_{-\infty}^{\infty} {\overline f}(x,t) \rho(x,t)  dx.
\ee

Define now the local mean velocity $u(x,t)$:
\be
u(x,t) \equiv \overline{\dot X} = {\< {\dot X} \delta(x-X) \>\over \rho},
\ee
and its root mean square counterpart $u'(x,t)$:
\be
{u'}^2 \equiv \overline{(\dot{X}-u)^2} =  \overline{\dot X^2}-u^2,
\ee
the mean kinetic energy of the particle is given by:
\be
\<E\> = \int_{-\infty}^{\infty} {m\over 2}\overline{\dot X^2}\rho dx= \int_{-\infty}^{\infty}  {m\over 2}(u^2 +{u'}^2)\rho dx.
\ee

Next we will obtain a continuity-like equation. The time derivative of the PDF gives:
\be
\der{\rho}{t} = \der{}{t} \<\delta(x-X(t)) \> = \<\dot{X} \der{}{X} \delta(x-X) \> = - \der{}{x} \<\dot{X}\delta(x-X) \>.
\ee
Using (8) we therefore obtain:
\be
\der{\rho}{t} = - \der{}{x}(\rho u).
\ee
For the ensemble mean momentum-like equation we take the time derivative of the momentum flux,
when we first use the definition of (6) for ensemble means:
\be
\der{}{t}(\rho u) =\der{}{t}\<\dot{X}\delta(x-X)\>
= \rho\overline{\ddot{X}} - \der{}{x}( \rho\overline{\dot X^2})
\ee
and second when using the continuity equation (12):
\be
\der{}{t}(\rho u) = \rho\(\der{u}{t}+ u\der{u}{x} \) - \der{}{x}(\rho u^2) = \rho{Du\over Dt} - \der{}{x}(\rho u^2),
\ee
where ${D\over Dt}\equiv (\der{}{t}+ u\der{}{x} )$ is the ensemble mean material time derivative. Equating the RHS of (13) and (14) we obtain
\be
{Du\over Dt} = \overline{\ddot{X}} - {1\over \rho}\der{}{x}\[\rho \overline{(\dot{X}-u)^2}\].
\ee
Using (3) and (9) we can now write:
\be
{Du\over Dt} = {\overline{F}\over m} - {1\over \rho}\der{}{x}(\rho {u'}^2).
\ee



A familiar straightforward application of this formulation is the ideal gas model (recall that the ensemble mean statistical properties are equivalent to the ones of $N$ co-existing, non-interacting, identical particles moving stochastically). In ideal gas the local mean of the stochastic force is assumed to be zero ($\overline{F} = 0$) and
\be
p_g = \rho m {u'}^2,
\ee
is the thermodynamic pressure satisfying:
\be
{p_g\over \rho} = m {u'}^2 = k_B T,
\ee
where $k_B$ is the Boltzmann constant and $T$ is the ideal gas temperature.
In this context, (12) is the hydrodynamic continuity equation where $u$ represents the hydrodynamic velocity. Equation (16) becomes the Euler fluid momentum equation (in the absence of external forces):
\be
m{Du\over Dt} =  - {1\over \rho}\der{p}{x}
\ee
and the mean energy of (10) becomes:
\be
\<E\> = \int_{-\infty}^{\infty} (K+I)\rho dx,
\ee
where $K = mu^2/2$ is the specific hydrodynamic kinetic energy and $I = m {u'}^2/2 =  k_B T/2$ is the specific internal energy.

Another example is the approach taken by Nikolic \cite{Nik} to describe the classical motion of a particle in vacuum as a statistical ensemble mean. In the classical limit, and in the absence of external forces, it is straightforward to show that Nikolic's nonlinear SE is equivalent to (12) together with the condition ${Du\over Dt} = 0$. Therefore, under this statistical approach to classical physics, (15) implies the relation:
$\rho\overline{\ddot{X}} = \der{}{x}\[\rho \overline{(\dot{X}-u)^2}\]$.

\section{Implications to quantum mechanics}

The real part of SE gives:
\be
\der{\rho}{t} =-\der{J}{x},
\ee
where the familiar density flux $J$ satisfies:
\be
J = {\rho \over m}\der{S}{x}.
\ee
Hence, we write (21) as the continuity equation (12) when define:
\be
u \equiv J/\rho = {1\over m}\der{S}{x} = {1\over m} \Re{\({\hat{p}\,\Psi \over \Psi}\)}.
\ee
Furthermore, from direct integration of SE we obtain (see also \cite{Yasue}):
\be
\<E\> = \int_{-\infty}^{\infty}-{\hbar^2\over 2m}\Psi^{*}\der{^2 \Psi}{x^2}dx =
\int_{-\infty}^{\infty} {m\over 2}(u^2 +{u'}^2)\rho dx = \int_{-\infty}^{\infty} (K+I)\rho dx ,
\ee
when $u'$ is defined as:
\be
u' = -{\hbar \over 2m}\der{}{x}\ln{\rho} = {1\over m} \Im{\({\hat{p}\,\Psi \over \Psi}\)}
\ee
(sometimes denoted as osmotic, or drift velocity \cite{Entropy,Cal04}), where $K ={mu^2\over 2}$ and $I = {m{u'}^2\over 2}$ can be regarded as kinetic and internal energy, respectively.

Taking now the spatial derivative of the imaginary part of SE, together with the definition of (23) we obtain the Madelung equation:
\be
m{Du\over Dt} = -\der{Q}{x},
\ee
where
\be
{Q}  = - {\hbar^2 \over 2m\sqrt{\rho}}{\der{^2\sqrt{\rho}}{x^2} }
\ee
is the Bohm potential \cite{Bohm,Dennis}.

Apart from (24), a different decomposition of $\langle E \rangle$ to $\langle K \rangle + \langle Q \rangle$ is well known in literature (see for instance \cite{Holl,Ori}), but in some cases \cite{HC,Entropy}, $\langle I \rangle$, which is proportional to the local Fisher information, can be helpful as well. Both decompositions are equally valid since, as shown in \cite{HC}, $\langle I \rangle = \langle Q \rangle$ (the latter identity {is shown explicitly in the Appendix}). Furthermore $I$, like $K$, but unlike $Q$, is a positive definite quantity.

Using the definition of (25) we can write the Bohm potential gradient as:
\be
-{1\over m}\der{Q}{x} = \({\hbar \over 2m}\)^2  {1 \over \rho} \der{^3\rho}{x^3} - {1\over \rho}\der{}{x}\(\rho {u'}^2\),
\ee
which is the 1D version of the decomposition suggested by Holland \cite{Holl} equations (3.10.11 - 3.10.12) and by Maddox and Bittner \cite{MadBit} (equation 38).
Substituting back in (26) we obtain:
\be
{Du\over Dt} =  \({\hbar \over 2m}\)^2  {1 \over \rho} \der{^3\rho}{x^3} - {1\over \rho}\der{}{x}\(\rho {u'}^2\).
\ee
Hence, equating the RHS of (16) and (29) gives an expression for the local mean stochastic quantum force:
\be
\overline{F} = {m}\({\hbar \over 2m}\)^2  {1 \over \rho} \der{^3\rho}{x^3}.
\ee
Equation sets (8,23); (9,25) and (3,30) point to a novel ensemble mean formulation of the time dependent SE.
This ensemble description may resonate with the previously suggested ensemble (or statistical) interpretation of quantum mechanics \cite{Sla29,Sch32,Bal70} or the more recent real ensemble interpretation \cite{Smo16}, but at this stage we keep the analysis general and unaffiliated with a specific interpretation.


It is straightforward to verify that the global mean of the stochastic quantum force is zero ($\<F\>=0$).
Information on its higher moments can be obtained when using (6) and (30) to write:
\be
\rho\overline{F} = \<F\delta(x-X)\> = {m}\({\hbar \over 2m}\)^2 \der{^3\rho}{x^3},
\ee
and implement the Fourier transform:
\be
{\tilde f}(q) = \int_{- \infty}^{\infty} f(x) e^{-iqx}dx,
\ee
so that
\be
{\tilde \rho} = \<e^{-iqX}\> = \sum_{k=0}^{\infty}{(-iq)^k \over k!}\<X^k\>.
\ee
The Fourier transform of (31) then yields:
\be
\sum_{k=0}^{\infty}{(-iq)^k \over k!}\<X^k F\> = -{1\over m}\({\hbar \over 2}\)^2 \sum_{n=0}^{\infty}{(-iq)^{n+3} \over n!}\<X^n\>,
\ee
implying $\<{F} \> = \<X {F} \> = \<X^2 {F} \> =0$. The second term shows that the mean work performed by the stochastic quantum force is zero, hence has zero contribution to the energy integral of (24).
Comparing in (34) the $k=3$ term at the LHS with the $n=0$ one at the RHS gives
\be
\<X^3 {F} \> = -{3\hbar^2\over 2m}.
\ee
For completeness, the general relation for higher moments is given by
\be
\<X^{k+3} {F} \> = -{1\over m}\({\hbar \over 2}\)^2 (k+1)(k+2)(k+3)\<X^k\>.
\ee

\section{Dynamic Gaussian example}

To obtain some familiarity with these results we consider as an example the dynamic Gaussian solution of the 1D SE \cite{Entropy,Tsekov12} for a free particle:
\be
\rho(x,t) =  {1\over \sigma(t) \sqrt{2\pi}}e^{-{x^2 \over 2\sigma^2(t)}}.
\ee
Substituting (37) in the continuity equation (12) yields:
\be
u(x,t) = x\der{\ln\sigma}{t},
\ee
where substitution of (37) and (38) back in (26) gives:
\be
\sigma\der{^2\sigma}{t^2} = \({\hbar \over 2m\sigma}\)^2 \hspace{0.25cm} \Longrightarrow  \hspace{0.25cm}
\sigma^2 = \sigma_0^2 + \({\hbar t\over 2m\sigma_0}\)^2.
\ee
Substitute (39) back in (38) and in (25) we then obtain
\be
u = \({\hbar \over 2m\sigma_0}\)^2{xt \over \sigma^2} =  {xt \over \({m\sigma_0^2\over \hbar}\)^2 +t^2}\, ; \hspace{0.25cm}
u' = {\hbar x \over 2m \sigma^2} = \({2m \sigma_0^2\over \hbar}\){u\over t}.
\ee
Hence while at short times $u \sim xt$ and $u' \sim x$, at long times $u \sim x/t$ and $u' \sim x/t^2$.
The particle's mean energy
\be
\<{m\over 2}{\dot X}^2\> = \int_{-\infty}^{\infty} {m\over 2}(u^2 +{u'}^2)\rho dx=\<K+I\> = {1\over 2}\({\hbar \over 2m\sigma_0}\)^2,
\ee
can be obtained either from direct integration of $\rho {m\over 2}(u^2 +{u'}^2)$, or from noting that the mean work of the stochastic force
is zero, $\<m{\ddot X} X\> = 0$. The latter then gives
\be
\<{m\over 2}{\dot X}^2\> = {m\over 4} \der{^2}{t^2}\<{X}^2\>.
\ee
Since $\<{X}^2\> = \sigma^2$ is the Gaussian variance distribution, (39) provides directly the particle energy.
To obtain the force partition we substitute (37) in (28):
\be
m{Du\over Dt} = -\der{Q}{x} = m\({\hbar \over 2m}\)^2 {x \over \sigma^4} = \overline{F} -{m\over \rho}\der{}{x}\(\rho {u'}^2\),
\ee
with
\be
 -{m\over \rho}\der{}{x}\(\rho {u'}^2\) = m{Du\over Dt}\[\({x \over \sigma}\)^2 -2\],
\ee
and
\be
\overline{F} =  m{Du\over Dt}\[3 - \({x \over \sigma}\)^2\].
\ee
Hence, when $|x| < \sqrt{2} \sigma$ the mean stochastic force per unit mass is larger than the mean acceleration of the particle
(i.e. $\overline{F}/m  >  {Du\over Dt}$) and vice-versa for $|x| > \sqrt{2} \sigma$. These terms become equal when $\der{p_g}{x} = 0$, for $\rho = 1/( \sqrt{2\pi}\sigma e)$.

It is interesting to compare the spreading of the Gaussian to the case of classical diffusion as well to the case of classical mechanics without determinism discussed by Nikolic \cite{Nik}. For classical diffusion it was shown by Heifetz et al. \cite{Entropy} that
$m{Du\over Dt} = +\der{Q}{x}$, where $D = {\hbar\over 2m}$ is the diffusion coefficient. Then the well known results of $\sigma = \sqrt{2Dt}$ and $u={x\over 2t}$ are recovered. In the second case  $m{Du\over Dt} =0$ so that $\overline{F} = {m\over \rho}\der{}{x}\(\rho {u'}^2\)$. Then $\sigma = \sigma_0 +\lambda t$ and
$u={\lambda x\over \sigma}$, where $\lambda$ is a non determined linear spreading coefficient.\\

We note as well that for the dynamic Gaussian solution of (37), equality (35) can be related to the Heisenberg inequality.
This can be shown when recalling that:
\be
 \<(X\dot X)^2 \>  = {\hbar^2\over 2m^2}  -{1\over 3}{d\over dt}{\<X^3 {\dot X} \>}.
\ee
Using the Cauchy-Schwarz inequality we obtain $\sqrt{\<{\dot X}^4 \> \< X^4\>} \ge \<(X\dot X)^2 \>$. Since $\<{\dot X}^4 \> = 2\<{\dot X}^2 \>^2$ and $\<{X}^4 \> = 2\<{X}^2 \>^2$ we obtain that $\sqrt{\<{\dot X}^4 \> \< X^4\>} = 2{\sqrt{\<{\dot X}^2 \>}\sqrt{\<{X}^2 \>}}$.
Defining now the time averaging operator as $\overline{f(t)}^t \equiv \lim_{T\rightarrow\infty}{1\over 2T}\int_{-T}^{T} f(t)dt$, then for Gaussian solutions
$\overline{{d\over dt}{\<X^3 {\dot X} \>}}^t  =0$, so that
\be
\overline{\sqrt{\<{\dot X}^2 \>}\sqrt{\<{X}^2 \>}}^t \ge {\hbar \over 2m},
\ee
which is inline with the more general results of \cite{Tsekov16} for the stochastic dynamics of a particle in vacuum.

\section{Discussion}

Up to now we have avoided explicit interpretational issues and simply compared the ensemble mean properties of identical classical particles, obeying a non-prescribed stochastic forcing, with the Madelung equations which are extracted directly from the Schr\"{o}dinger equation. The comparison revealed equation (30) which can be regarded as an expression for the local mean of the stochastic quantum force. A natural question to be asked now is what can be the origin of this stochastic force.

The second term of the RHS of (29) has a clear analogy to the ideal gas pressure gradient force of (17).
Writing analogously (30) as:
\be
\overline{F} = -{1 \over \rho} \der{p_v}{x},
\ee
results in
\be
p_v = -{1\over m}\({\hbar \over 2}\)^2\der{^2{\rho}}{x^2},
\ee
which has been previously interpreted as the pressure resulting from the vacuum fluctuations acting on the particle \cite{Tse16}. Then (25), (27) and (49) yield:
\be
Q = I +{p_v \over \rho},
\ee
suggesting the Bohm potential $Q$ to be the enthalpy associated with the vacuum fluctuations. For an inertial observer the temperature of the vacuum is zero, hence this enthalpy is equal to the Gibbs energy which in its turn is equal to the chemical potential. Therefore, the quantum Bohm potential can be understood as the energy required to put a particle in a bath of fluctuating vacuum at constant entropy and volume. For the dynamic Gaussian example the pressure distribution can be obtained when substituting (37) in (43) and (49):
\be
p =  p_g +p_v =   {\rho \over m}\({\hbar \over 2 \sigma}\)^2 ; \,\,\,
p_g = {\rho \over m}\({\hbar x \over 2 \sigma^2}\)^2 ; \,\,\,
p_v = {\rho \over m}\({\hbar \over 2 \sigma}\)^2\[ 1- \({x\over \sigma}\)^2\],
\ee
thus the vacuum pressure is positive (negative) for values of $|x|$ which are below (above) the standard deviation $\sigma$.


We shall now conclude with two remarks. First, it should be reminded that the analysis presented here has been performed for the 1D case, which
is relatively straightforward and contains the essence of the physics involved. The generalization to 3D dynamics is somewhat cumbersome and therefore it was decided not to present it in this short paper. Second, the actual stochastic law describing the force whose local mean yields Eq. (30) is yet to be found. We hope that in future work we will be able to better characterize the stochastic motion arising from this model, and in particular, to write explicitly the resulting equations of motion. These might shed further light on the origin of quantum stochasticity.\\

\section*{APPENDIX}

In what follows we show explicitly that $\<Q\>  = \<I\>$.

\setcounter{equation}{0}
\renewcommand{\theequation}{A.\arabic{equation}}
Using (27) we write:
\be
\<Q\>  = \int_{-\infty}^{\infty} \rho Q dx = -{\hbar^2 \over 2m} \int_{-\infty}^{\infty}\sqrt{\rho}{\der{^2\sqrt{\rho}}{x^2} }dx.
\ee
Therefore:
\be
\<Q\>  = -{\hbar^2 \over 2m} \int_{-\infty}^{\infty}\der{}{x}\(\sqrt{\rho}{\der{\sqrt{\rho}}{x}}\)dx +
{\hbar^2 \over 2m} \int_{-\infty}^{\infty}\(\der{\sqrt{\rho}}{x}\)^2dx.
\ee
The first integral in the RHS vanishes when assuming that $\rho \rightarrow 0$ smoothly as $x \rightarrow \pm \infty$.
Hence:
\be
\<Q\>  = {\hbar^2 \over 2m} \int_{-\infty}^{\infty}\rho \({1\over\sqrt{\rho}} \der{\sqrt{\rho}}{x}\)^2dx =
{\hbar^2 \over 2m} \int_{-\infty}^{\infty}\rho \({1\over2} \der{\ln \rho}{x}\)^2dx,
\ee
so that:
\be
\<Q\>  = \int_{-\infty}^{\infty}\rho \[{\hbar^2 \over 8m}\(\der{\ln \rho}{x}\)^2\]dx \equiv \<I\>,
\ee
where
\be
I = {\hbar^2 \over 8m}\(\der{\ln \rho}{x}\)^2 = {m \over 2}{u'}^2,
\ee
and $u'$ is defined in (25).

\begin{acknowledgments}

The authors wish to thank two anonymous reviewers for their elaborated reports which helped to improve the manuscript. E.H. is grateful for an illuminating discussion with Shay Zucker. E.C. was supported by ERC AdG NLST.

\end{acknowledgments}
\goodbreak

\end{document}